# Low field anisotropic colossal magnetoresistance in $Sm_{0.53}Sr_{0.47}MnO_3$ thin films

Manoj K. Srivastava,[1, 3] M. P. Singh,[2*] Amarjeet Kaur,[3] F. S. Razavi,[2] and H. K. Singh[1#]

[1]*National Physical Laboratory (CSIR), Dr. K. S. Krishnan Marg, New Delhi-110012, India*
[2]*Department of Physics, Brock University, St. Catharines, Ontario, L2S 3A1, Canada*
[3]*Department of Physics and Astrophysics University of Delhi, Delhi-110007, India*

**Abstract**

$Sm_{0.53}Sr_{0.47}MnO_3$ (SSMO) thin films (thicknesses ~200 nm) were deposited by on-axis dc magnetron sputtering on the single crystal LSAT (001) substrates. These films are oriented along the out of plane c-direction. The ferromagnetic and insulator-metal transition occurs at $T_C$~ 96 K and $T_{IM}$~ 91 K, respectively. The magnetization easy axis is observed to lie in the plane of the film while the magnetic hard axis is found to be along the normal to this. The magnetotransport of the SSMO films, which was measured as a function of angle (θ) between the magnetic field (**H**) and plane of the film, shows colossal anisotropy. Magnetoresistance (MR) decreases drastically as θ increases from 0° (**H**//easy axis) to 90° (**H**//hard axis). The out-of-plane anisotropic MR (AMR) is as high as 88 % at H=3.6 kOe and 78 K. The colossal anisotropy has been explained in terms of the magnetic anisotropies at play and the magnetic domain motion in applied magnetic field.

[#]Corresponding author: hks65@nplindia.org

[*]mangala.singh@brock.ca

**Introduction**

The rich electronic phase diagram of the doped rare earth manganites ($RE_{1-x}AE_xMnO_3$, where RE = La, Nd, Sm, etc. and AE = Sr, Ca, Ba, etc.) is a natural consequence of the strong coupling between the mutually interacting spin, charge, orbital and lattice degrees of freedom. The phase diagram becomes more complex with the appearance of instabilities and phase interdiffusion causing multicricalilty with increasing size mismatch between the RE and AE cations. $Sm_{1-x}Sr_xMnO_3$ by virtue of being close to the charge/orbital order (CO/OO) instability possesses intrinsic phase instabilities and multicritical points. This low bandwidth (BW) compound shows variety of ground states such as (i) ferromagnetic metal (FMM) at $0.3<x\leq0.52$ (ii) antiferromagnetic insulator (AFMI) for $x>0.52$ and (iii) for $0.4<x\leq0.6$, the charge ordering (CO) occurs with the ordering temperature ($T_{CO}$) increasing from ~140 K to 250 K as x increases in the above range.[1-3] The competition between the FM, CO/OO, and AFM states becomes dominant near half doping ($x=0.5$), and tricritical peculiarities are observed in the $x$-dependent electronic phase diagram making $Sm_{1-x}Sr_xMnO_3$ an interesting candidate to study phase competition and related phenomena.[2,5] Low field colossal magnetoresistance (LF-CMR) is observed at all the compositions that corresponds to the FMM ground state in the range $0.3<x\leq0.52$. A first order transition from paramagnetic insulating (PMI) to the FMM state is observed in $Sm_{1-x}Sr_xMnO_3$ ($x\sim0.45-0.50$), which is technologically important for applications such as low temperature low field magnetoresistive devices, bolometers and large magnetocaloric effect (MCE) applications.[4,5] The low BW and strong phase coexistence/phase separation also result in metamagnetism that provides intrinsic fragility to the composition-temperature ($x$-T) phase diagram vis-a-vis external perturbations. Consequently, even mild external perturbation such as electromagnetic field, pressure, lattice strain provided by the substrate, etc. could dramatically modify their physical properties.[1-3] Extensive investigations have been reported on $Sm_{1-x}Sr_xMnO_3$ in poly- and single- crystalline bulk forms.[2-7] However, thin films of this compound, either polycrystalline or epitaxial have not been studied in much detail.[8-10]

Another important property of doped rare earth manganites is anisotropic magneto resistance (AMR) due to anisotropic magneto-crystalline nature that results in the dependence of resistivity on the angle between the applied magnetic field (**H**) and the direction of magnetic easy axis or transport current (**J**).[11-21] The magnetic anisotropy is generally defined as the energy required to rotate the magnetization direction from the easy to the hard axis.[20] Magnetocrystalline anisotropy originates from the interaction between the electronic spin and orbital degrees of freedom. The electronic orbitals are linked to the crystallographic structure and their interaction with the spins aligns the latter preferentially along well defined crystallographic axes. Therefore, there are directions in space, (generally referred to as the easy axis) along which a magnetic material is easier to magnetize in than in other ones. However, polycrystalline samples without a preferred granular orientation do not possess any magnetocrystalline anisotropy. But, an overall isotropic behavior concerning the energy needed to magnetize it along an arbitrary direction is only given for a spherical shape. For non-spherical shapes there are one or more specific directions, solely caused by the shape which represents easy magnetization axes. This anisotropy is known as shape anisotropy.[20] The direction of magnetization is determined by the competing magnetocrystalline and shape anisotropies. The constant characterizing magneto crystalline anisotropy is found to be smaller than that characterizing the shape anisotropy and hence the later dominates the former which results in an in-plane magnetization in thin films. In low dimensional systems such as thin films, the anisotropies are affected further by the broken symmetry at the interfaces and hence additional contributions that are forbidden in three dimensional cases arise. In manganite thin films the non-spherical charge distribution around Mn ions gets further modified by substrate induced strain and lattice defects.[9,10] Apart from the strain and hence the film thickness other factors such as the structural defects, spin disorder, nature of the magnetic ground state, phase coexistence, etc. are also expected to play a crucial role. The occurrence and nature of the low field AMR in low bandwidth manganites with strong phase coexistence such as SSMO has not been given much attention. In the present work we have studied the variation of LF-MR as a function of the angle ($\theta$) between the applied magnetic field (**H**) and current applied (**J**) in the plane of the film. Huge anisotropy is observed in magnetotransport properties. AMR as high as ~85 % is observed at T=81 K and H=3.6 kOe. The low field MR is observed to decreases from ~80 % to 6 % as angle ($\theta$) between **J** and **H** increases from 0° to 90°.

**Experimental Details**

$Sm_{0.53}Sr_{0.47}MnO_3$ (SSMO) thin films (thicknesses ~200 nm) were deposited by on-axis dc magnetron sputtering on the single crystal LSAT (001) substrates from 2 inch diameter target prepared by solid state reaction route. Before deposition of the films, the chamber was cleaned by attaining a vacuum of up to ~$10^{-6}$ torr and then flushed twice by an Ar (80%) +$O_2$ (20%) gas mixture. The substrate temperature was maintained at 800 °C at a dynamic pressure of 200 mtorr of Ar -$O_2$ mixture. After deposition, the films were kept at 800 °C for half an hour at 1.0 torr gas pressure and then slowly cooled down to room temperature. The films were then annealed in flowing oxygen at 900 °C for about 12 hrs. The structural characterization was performed by X-ray diffraction (XRD, θ-2θ and ω-2θ scans). The cationic composition was studied by energy dispersive spectroscopy (EDS) attached to scanning electron microscope. The temperature and magnetic field dependent magnetization was measured by a commercial (Quantum Design) MPMS at 100 Oe magnetic field applied parallel as well as perpendicular to the film surface. The temperature and magnetic field dependent electrical resistivity was measured by the standard four probe technique. The temperature and magnetic field dependent resistivity of the SSMO films was measured as a function of the angle (θ) between **H** and the direction of the current **J**. The current, which is always in the plane of the film, is applied along the larger dimension of the film. Since the current is fixed in the plane of the film along the magnetization easy axis lays, therefore **H** ∥ **J** and **H** ⊥ **J** would mean that **H** is parallel and perpendicular to the magnetization easy axis, respectively. From these measurements angular dependence of magnetoresistance and the AMR was calculated. The AMR measured in the present study is defined as AMR = ($\rho^{in}_{\parallel}$ - $\rho^{out}_{\perp}$)*100/$\rho_{av}$. $\rho^{in}_{\parallel}$ is the resistivity of the film for **H** located in the plane of the film and oriented parallel to **J**. $\rho^{out}_{\perp}$ represents the resistivity measured when **H** is applied perpendicular to both the plane of the film (magnetic easy axis) and the current **J**. The average resistivity is defined to be $\rho_{av}$ = $\rho^{in}_{\parallel}$/3 + 2$\rho^{out}_{\perp}$/3.

**Results and Discussion**

The thickness of the films was measured by a profilometer (AMBIOS technology, XP200) on step structured films prepared simultaneously and was found to be ≈200 nm. The cationic composition of the films was verified by EDS from several places and the average composition was found to be in excellent agreement with that of the target. The occurrence of only the (00ℓ) diffraction maxima in the θ-2θ scan of SSMO shows highly oriented growth (main panel of Fig. 1). This is further confirmed by the ω-2θ scan of the (002) peak (inset of Fig. 1). The full width at half maximum (FWHM) of the (002) peak is estimated to be 0.11°. This shows very good crystallinity in the present film. The out of plane lattice parameter ($a_c$) estimated from the XRD data is 3.833 Å, which is in good agreement with the corresponding value for the single crystals of similar composition.[2]

The temperature dependent zero field cooled (ZFC) and field cooled (FC) magnetization data (M(T)) taken at H=100 Oe applied parallel to the film surface is shown in Fig. 2. The PM-FM transition temperature ($T_C$) is ~96 K. Further, as we lower the temperature $M_{ZFC}(T)$ shows a cusp like feature at $T_P \approx 53$ K, and then drops sharply below this point. In the $M_{FC}(T)$ curves, the cusp is shifted to lower temperature ($T_P$~36 K) and the sharpness of the magnetization drop at $T<T_P$ is reduced considerably. The $M_{ZFC}(T)$ and $M_{FC}(T)$ curves show divergence at $T<T_C$ that could be considered as a signature of the cluster glass state. Generally, ZFC and FC divergence, the cusp-like behavior in the ZFC magnetization and the drop in ZFC magnetization below $T_P$ are the typical signatures of metamagnetic systems such as spin glass (SG)/cluster glass (CG).[22] Since the difference between the ZFC and FC, which starts at temperature just below $T_C$, is large and the magnetization drop is seen even in the FC curve the possibility of the occurrence of a true SG system in the present case is ruled out. Hence we believe that the magnetic order is spin cluster glass (CG) type. The M-H data (taken at 10 K) measured with **H** oriented parallel (θ=0°) and perpendicular (θ=90°) to the film surface is presented in the inset of Fig. 2. When θ = 0°, the saturation magnetization ($M_{S\parallel}$) ≈ 467 emu/cm$^3$ is observed at a saturation field of $H_{S\parallel}$ ≈ 10.1 kOe and remnant magnetization is ($M_{r\parallel}$) ≈ 200 emu/cm$^3$. When **H** is normal to the film surface (θ= 90°) the saturation moment reduces by 15 % ($M_{S\perp}$ ≈ 396 emu/cm$^3$) and the saturation field increases slightly to $H_{S\perp}$ ≈ 10.9 kOe. In contrast the remnant magnetization is reduced ($M_{r\perp}$ ≈ 84 emu/cm$^3$). These results suggest that the easy axis of the magnetization lies in the plane of the film.

Fig. 3 shows the representative temperature dependence of resistivity (ρ-T) measured in the range 4.2-300 K using the protocol of slow cooling and subsequent heating in zero magnetic field. On cooling from the room temperature, the resistivity gradually grows until an insulator-metal transition at $T_{IM}$ ≈ 80 K, which is accompanied by unusually huge and sharp drop in resistivity. During the heating cycle, the resistivity grows gradually out of the low temperature flat region and the IMT occurs at ~91 K. The resistivity drops by more than three orders of magnitude within a very small temperature range. Such drop in the resistivity is attributed to bicriticality in the low bandwidth manganites.[2] The sharp IMT is accompanied by a irreversibility (hysteresis) in the resistivity, which is observed in temperature range 50 -100 K.

Such behaviour indicates irreversible thermal changes in the sample resistivity and resembles thermo-remnant magnetization in spin glasses, which are well known for their nonequilibrium slow dynamics such as long-time relaxation, aging, and memory behaviors.[22]

It is previously reported that there is a coexistence of AFM-CO/OO insulating and FM metallic phases in SSMO.[1-7] The length scale and behavior of these coexisting phases is strongly temperature dependent. In the vicinity of the room temperature, the SSMO is PMI. On lowering the temperature from room temperature the fraction of the AFM-CO/OO insulating phase, which is short range in nature, is enhanced.[1,2] The much sharper rise in resistivity just above $T_{IM}$ is evidence of this. Just above the $T_C/T_{IM}$ the fraction of the AFM-CO/OO insulating phase raises rapidly resulting in a sharper rise in the resistivity.[1,2] At $T_C/T_{IM}$ the AFM-CO/OO insulator and FMM coexistence is delicately balanced and possesses bicriticality. In such a situation any external perturbation such as temperature and magnetic field can easily destroy the unfavorable phase. Consequently, the AFM-CO/OO insulator phase is suddenly removed (the FMM phase appears) when the temperature is even slightly lower than the $T_C$. This explains the sharp drop in resistivity observed in the vicinity of $T_C$. However, even at $T < T_C$, the short range AFM-CO/OO insulating clusters are present in the FMM background and may cause the occurrence of cluster glass (CG) like metamagnetic configuration. The irreversibility in the ZFC-FC M(T) curves resulting in the formation of the spin CG as explained earlier is as an evidence of AFM-CO/OO clusters in the FMM matrix. Occurrence of huge MR even at low magnetic fields (discussed below) shows the soft nature of the CG as well as the fragility of the CO state with respect to external magnetic field. The observed glass like behaviour transport (hysteresis in the ρ-T data) is commonly attributed to the slow evolution of the phase conversions among competing phases coexisting in the material as a result of phase separation. The hysteresis in the resistivity and concomitant difference in the IMTs could be understood in terms of the cluster glass concept. During the cooling cycle the spin CG mimics a liquid like behavior where in the carrier scattering by disordered spins is stronger. This causes higher resistivity and the lower $T_{IM}$. The carrier scattering is strongly reduced due to induced spin order when magnetic field is applied. When cooled below the freezing temperature ($T<T_f$) the CG is frozen and in this frozen state the carrier scattering is considerably suppressed resulting in lower resistivity and higher $T_{IM}$ during the heating cycle.

We measured the anisotropic magnetotransport properties as function the angle (θ) between the in-plane transport current (**J**) or the magnetic easy axis and the magnetic field (H). The applied current was always in the plane of the film. The angle (θ) was varied from 0º (H ∥ **J**) to 90°(H ⊥ **J**). The temperature dependence of the in-plane (θ = 0°) MR, out-of-plane MR (θ = 90°) and the AMR is plotted in the inset of Fig. 3. The peak magnitude of the AMR calculated using ($\rho^{in}_{\parallel}$ - $\rho^{out}_{\perp}$)*100/$\rho_{av}$ is found to be ~88 % at H=3.6 kOe. Fig. 4 (a - d) shows a representative plot of the magnetic field driven isothermal resistance change (78 K) for different angles (θ = 0º, 30º, 60º and 90º) between the applied magnetic field H and **J** flowing through the film. As seen in the plot 4a, when H is parallel to **J** or magnetic easy axis (θ = 0°) the resistance drops sharply on application of the magnetic field, resulting in low field colossal magnetoresistance (LF-CMR) ~99 % at H = 3.6 kOe. On reversing the magnetic field the resistance grows again but the virgin resistivity could not be attained even at H = 0. Instead the maximum resistance was attained at H ≈ -0.3 kOe. In the subsequent cycles of the magnetic field produced an irreversible growth and decay of resistance that resulted in a hysteresis. The LF-CMR corresponding to the resistance variation at θ = 0° is shown in the inset of Fig. 4a. The LF-CMR ~99 % observed during the virgin cycle is reduced to ~80 % (both at 3.6 kOe) during the subsequent ones. At θ = 30° the isothermal resistance as well as LFMR show nearly similar behavior (Fig. 4b). However, the magnitude of LF-CMR (at 3.6 kOe) is reduced during both the virgin (~90 %) as well as subsequent cycles (~70 %). As θ is increased further, isothermal resistance decay as a function of the magnetic field is drastically reduced and the observed hysteresis disappears. To demonstrate this we present the isothermal resistance and LFMR data measured at θ = 60° and 90° in Fig. 4c and 4d, respectively. At θ = 60° no hysteresis is seen and the maximum LFMR measured at H = 3.6 kOe in the virgin and subsequent cycles decreases to ~45 % and ~30 % respectively. When H becomes parallel to magnetic hard axis, that is perpendicular to **J** (θ = 90°) the maximum LFMR measured at H = 3.6 kOe in the virgin and subsequent cycles decreases to ~21 % and ~7 % respectively. The variation of the LFMR measured at H=3 kOe and 80 K as function of the angle between the applied in-plane transport current (**J**) and the magnetic field is plotted in Fig. 5. Our results demonstrate that the SSMO thin film on LSAT shows huge out-of-plane (OP) anisotropy in the electrical transport properties.

The field dependence of LFMR and the variation of their magnitude could be understood in terms of the magnetic anisotropies at play and the magnetic domain motion in applied magnetic field. When the magnetic field is in the plane of the film, because of the in-plane easy axis, due to the strong in plane coupling of magnetization vector (**M**) with applied **H** results in maximum magnetic domains alignment along the field direction and resistance is lowered. When the magnetic field is applied perpendicular to the plane (out-of-plane), that is, along the magnetic hard axis, much larger magnetic field is required to rotate and align all the spins along the magnetic hard axis (**M** is weakly coupled to **H**). This difference in the in-plane and out-of-plane LFMR behaviour of the present films can also be understood in terms of the different types of anisotropies, such as, magneto crystalline, shape and surface anisotropy that come into play in thin films. If $\theta$ be the angle between **M** vector and the out-of-plane direction, then the stray energy density in case of a thin film is given by $E_{stray} = K_0 + K^V_{shape}\sin^2\theta$, where $K^V_{shape}$ is proportional to $-M^2 < 0$. Thus at $\theta = 90°$ the stray energy will be minimum and hence the shape anisotropy would favor a magnetization direction parallel to the film surface, that is, the easy axis lies in the plane of the film. Generally, the shape anisotropy dominates over the magneto crystalline anisotropy, which results in an in-plane magnetization for thin film systems. In case of low BW manganites such as SSMO around half doping ($x \sim 0.5$), the presence of AFM-COI inhomogeneities over a wide range around $T_C$ as well as in the FMM causes strong phase fluctuations and small external perturbations such as magnetic field, temperature, etc. can easily destabilize one phase at the cost of the other. The competing phases (FMM and AFM-COI) are pinned strongly into the plane of the film by the shape anisotropy, that is, the nature of the magnetic domains is rendered nearly two dimensional. As described earlier, the strong FMM-AFMI-CO competition results in formation of a CG like glassy phase. When H is in the plane of the films the magnetization vectors of this glassy CG phase is easily switched along the direction of the field resulting in a drastic decrease in the resistivity of the films. As the angle between **H** and the out-of-plane direction (c-axis) increases from $\theta = 0°$ towards $90°$, due to the strong inplane domain pinning lesser and lesser fraction of domains are switched along the direction of the magnetic field. This could be the reason for the much smaller change observed in case of **H** perpendicular to the plane of the films. Here we would like to mention that at $\theta \neq 0°$, non-negligible contribution from the Lorentz force could also arise.

## Conclusions

In summary, we have studied the out-of-plane magnetic and transport properties of oriented $Sm_{0.53}Sr_{0.47}MnO_3$ thin films grown on LSAT (001) single crystal substrates. Huge anisotropy is observed in the electrical transport when the current that is always in the plane of the film is normal to the applied magnetic field. Anisotropic MR as high as $\approx$-85 % is observed at T=81 K and a small magnetic field H=3.6 kOe. The isothermal resistivity (hence MR) measured as a function of the magnetic field shows hysteresis when the angle between the current and the magnetic field is smaller than ≤50°. The observed huge anisotropic MR has been discussed in terms of the effects caused by (a) the magnetic domain motion in applied magnetic field, (b) different types of anisotropies, such as, magneto crystalline, shape and surface anisotropy that come into play in thin films, and (c) the strong phase competition between FMM and AFM-COI phases.

## Acknowledgments

MKS is thankful to CSIR, New Delhi for a senior research fellowship. Support and encouragement from DNPL is thankfully acknowledged. At St Catharines, this work was supported by NSERC (Canada), CFI, Ontario Ministry of Research and Innovation (MRI), and the Brock University.

## References

1. Y. Tokura, Rep. Prog. Phys. **69**, 797 (2006).
2. Y. Tomioka, H. Hiraka, Y. Endoh, and Y. Tokura, Phys. Rev. B **74**, 104420 (2006).
3. C. Martin, A. Maignan, M. Hervieu, and B. Raveau, Phys. Rev. B **60**, 12191 (1999).
4. A. Rebello and R. Mahendiran, Appl. Phys. Lett. **93**, 232501 (2008).
5. M. Egilmez, K. H. Chow, J. Jung, and Z. Salman, Appl. Phys. Lett. **90**, 162508 (2007).
6. P. Sarkar, P. Mandal, and P. Choudhury, Appl. Phys. Lett. **92**, 182506 (2008).
7. M. Egilmez, K. H. Chow, J. Jung, I. Fan, A. I. Mansour, and Z. Salman, Appl. Phys. Lett. **92**, 132505 (2008).
8. H. Oshima, K. Miyano, Y. Konishi, M. Kawasaki, and Y. Tokura, Appl. Phys. Lett. **75**, 1473 (1999).
9. M. Kasai, H. Kuwahara, Y. Tomioka, and Y. Tokura, J. Appl. Phys. **80**, 6894 (1996).
10. W. Prellier, Ph Lecoeur, and B. Mercey, J Phys. Condens. Matter **13**, R915 (2001).
11. A.-M. Haghiri-Gosnet, and J.-P. Renard, J. Phys. D: Appl. Phys. **36** R127 (2003).

12. P. A. Stampe, H. P. Kunkel, Z. Wang, and G. Williams, Phys. Rev. **B52** 335 (1995).
13. Y. Suzuki, H. Y. Hwang, S. W. Cheong, and R. B. van Dover, Appl. Phys. Lett. **71,** 140 (1997).
14. H. S. Wang and Q. Li, Appl. Phys. Lett. **73,** 2360 (1998).
15. H. S. Wang, Qi Li, Kai Liu and C. L. Chien, Appl. Phys. Lett. **74,** 2212 (1999).
16. J. O'Donnell, J. N. Eckstein, and M. S. Rzchowski, Appl. Phys. Lett. **76**, 218 (2000).
17. M. Ziese, Phys. Rev. B **62**, 1044 (2000).
18. R. Patterson, C. Ozeroff, K. H. Chow, and J. Jung, Appl. Phys. Lett. **88**, 172509 (2006).
19. M. Egilmez, R. Patterson, K. H. Chow, and J. Jung, Appl. Phys. Lett. **90**, 232506 (2007).
20. M. Gezlaff, Fundametals of Magnetism, Springer-Verlag Berlin Heidelberg (2008).
21. G. Singh-Bhalla, S. Selcuk, T. Dhakal, A. Biswas, and A. F. Hebard, Phys. Rev. Lett. **102**, 077205 (2009).
22. J. A. Mydosh, Spin Glasses: An Experimental Introduction, 2nd ed., Taylor & Francis, London (1993).

**Figure Captions**

1. X-ray Diffraction pattern of SSMO film grown on LSAT (001) substrate. The inset shows the $\omega$-$2\theta$ scan of the (002) reflection.
2. Temperature dependent zero field cooled (ZFC) and field cooled (FC) magnetization (H@ 100 Oe) of SSMO film. The inset shows in-plane and out-of-plane M-H loops of SSMO film measured at 10 K.
3. Temperature dependent resistivity measured in cooling and heating cycles. Inset shows the variation of AMR and MR with temperature when **J** is parallel to **H** ($\theta$=0°) and perpendicular to it ($\theta$=90°).
4. Magnetic field dependence of resistance (at T=78 K) measured at different angles between the current and the magnetic field (a) $\theta$=0°, (b) $\theta$=30°, (c) $\theta$=60°, and (d) $\theta$=90°. The respective insets show the corresponding variation of MR.
5. The variation of the LFMR measured at H=3 kOe and T=80 K as function of the angle between the applied transport current and the magnetic field.

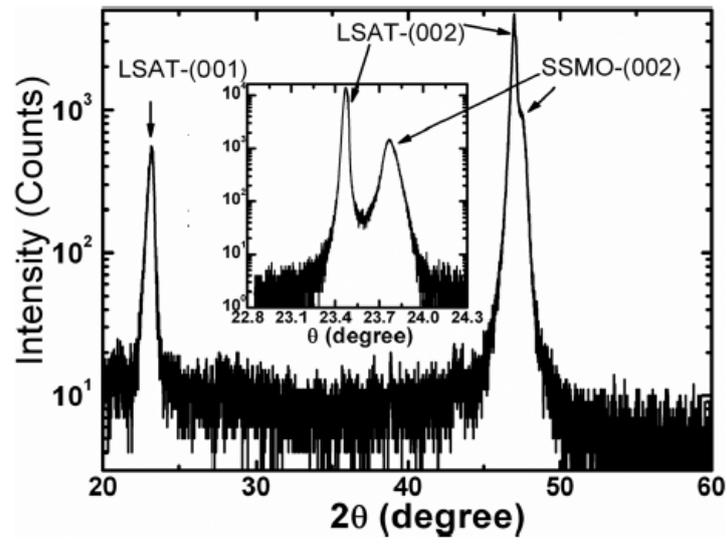

Fig. 1

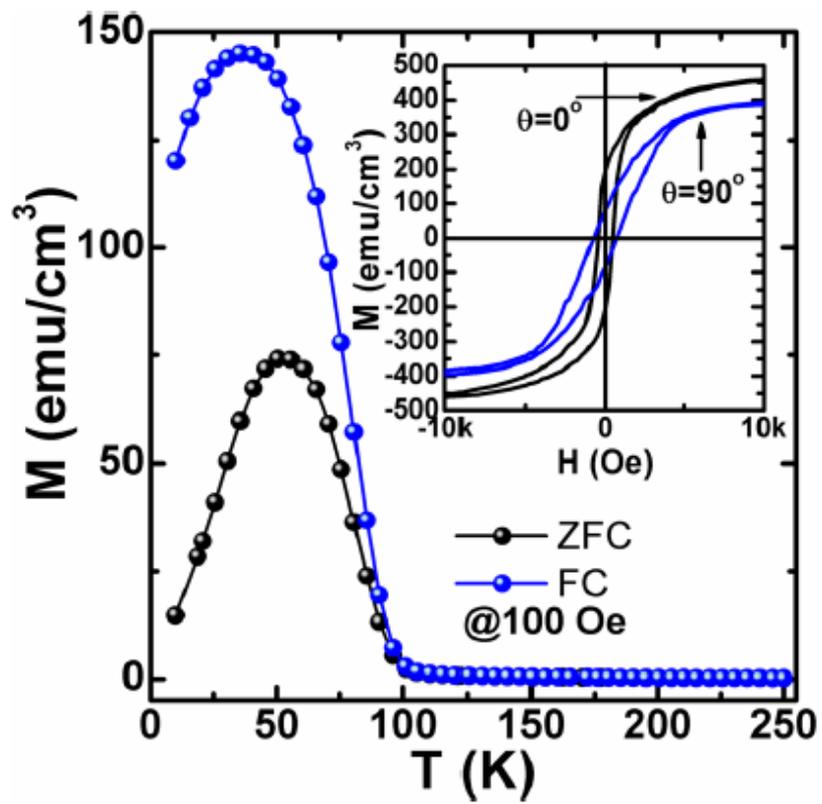

Fig. 2

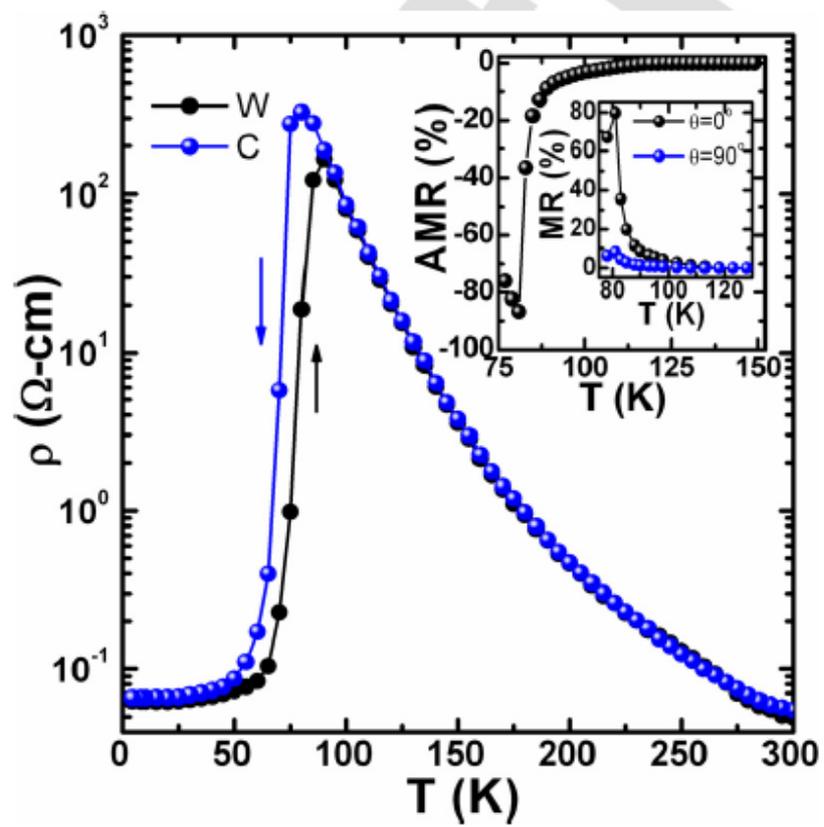

Fig. 3

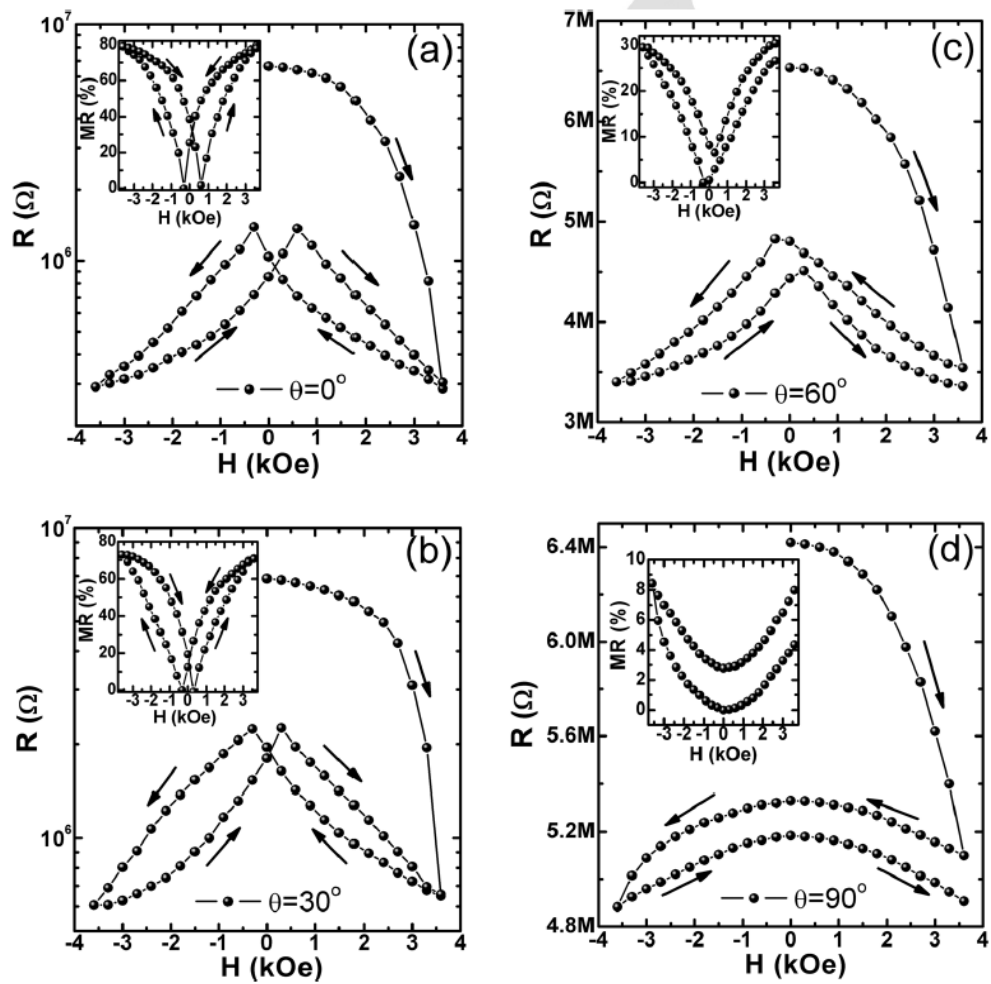

Fig. 4

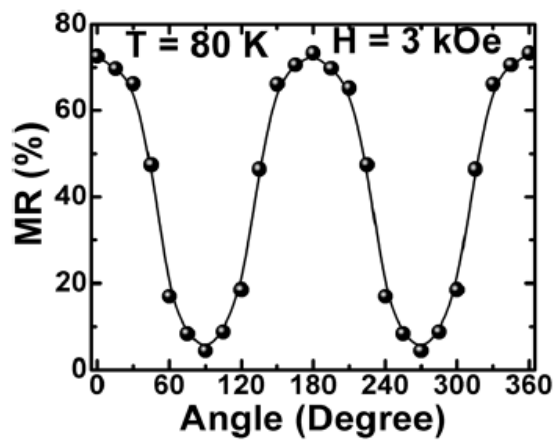

Fig. 5